\documentclass[a4paper,11pt]{article}
\usepackage{pos}
\newcommand{\y}{Y(4260)}

\newcommand{\z}{Z_c(3900)}
\newcommand{\x}{X(3872)}
\newcommand{\psit}{\psi_2(3823)}
\newcommand{\pp}{\pi^+\pi^-}
\newcommand{\pip}{\pi^+}

\newcommand{\kk}{K^+K^-}

\newcommand{\EE}{e^+e^-}

\newcommand{\GG}{\gamma\gamma}

\newcommand{\psip}{\psi(2S)}

\newcommand{\jpsi}{J/\psi}
\newcommand{\piz}{\pi^0}
\newcommand{\chico}{\chi_{c1}}
\newcommand{\chict}{\chi_{c2}}

\newcommand{\ppjpsi}{\pi^+\pi^-J/\psi}
\title{Spectroscopy at $\EE$ colliders}
\author*[a]{Zhiqing Liu (on behalf of BESIII Collaboration)}
\affiliation[a]{Shandong University,\\
  Binhai Road No. 72, Qingdao, China}

\emailAdd{z.liu@sdu.edu.cn}
\abstract{Exotic hadron states beyond the conventional $qqq$ baryon and $q\bar{q}$ meson configurations are well expected from QCD. Having been successful in the past decades, the heavy quarkonium sector is proved to be an ideal place for the study of exotic hadrons. In this talk, I briefly review the recent progress on exotic heavy quarkonium-like state at $\EE$ machines, including BESIII, Belle/Belle II etc.}

\FullConference{21st Conference on Flavor Physics and CP Violation (FPCP 2023)\\
 29 May - 2 June 2023\\
 Lyon, France\\}

\begin{document}
\maketitle
%. 1. introductin
\section{Introduction}
In 1964, with the birth of the famous Quark Model~\cite{quarkmodel}, people believed that hadrons are composite particles
and quarks are elementary. Many experimental observations later support that hadrons are built of three quarks 
$qqq$ (so called baryons) or quark anti-quark pairs $q\bar{q}$ (so called mesons)~\cite{pdg}.
The Quark Model successfully describes
the structure of most hadrons observed in experiments, and are even very powerful today. 
However, hadrons are bound states of quarks via the strong force. The fundamental theory for the strong force is QCD, 
which allows the existence of more color neutral combinations, such as multi-quark states (tetraquark, pentaquark, ...), 
quark-gluon hybrid states, glueballs etc. All the new combinations beyond the traditional baryon/meson picture
are referred as exotic states.

The existence of exotic states trigger big interests to hunt them in experiment. In the past two decades, 
lots of new particles (also called XYZ particles)
are discovered in the heavy quarkonium energy region~\cite{RMP}, and are considered good candidates for exotic hadrons.
\section{$\x$ and $\psit$ studies}
\begin{figure}[h]
\begin{center}
\includegraphics[height=2.0in]{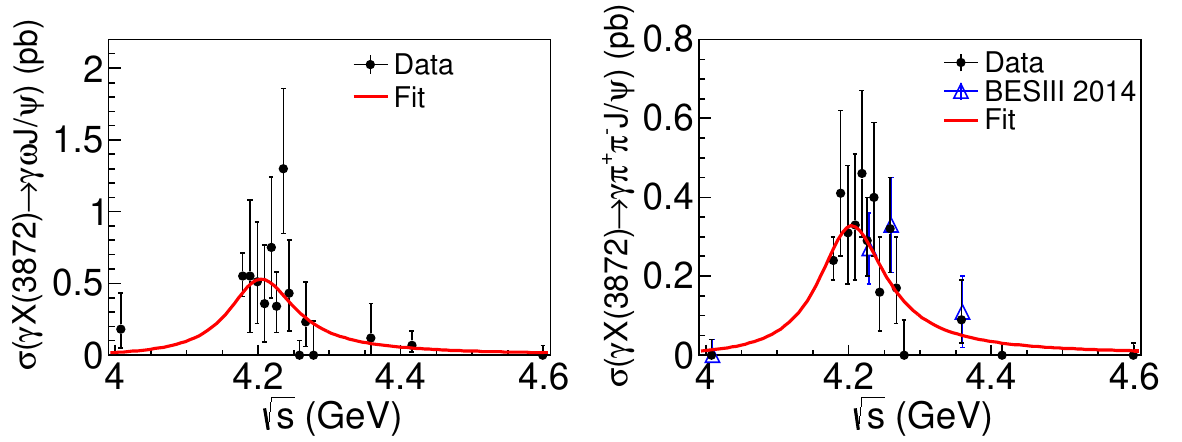}
\caption{The measured $\EE\to\gamma\x$ $\sqrt{s}$-dependent production cross section in $\omega\jpsi$
mode (left) and $\ppjpsi$ mode (right) at BESIII~\cite{bes3-wjpsi}.}
\label{xs-x3872}
\end{center}
\end{figure}

The $\x$ is the first charmonium-like state discovered by the Belle experiment in 2003~\cite{x3872}. By far, it's still one of the most
studied candidate. The BESIII experiment firstly observed it in the $\EE\to\gamma\x$ process~\cite{bes3-x3872}. 
With much more data (11.6~fb$^{-1}$),
BESIII is able to reconstruct about 90 signal candidates in the $\x\to\ppjpsi$ decay. 
With these events, we further study the $\sqrt{s}$-dependent production cross section of $\EE\to\gamma\x$ in two decay channels ($\omega\jpsi$ and $\ppjpsi$)~\cite{bes3-wjpsi}. A clear enhancement
is visible near $4.2$~GeV in the cross section line shape, as shown in Fig.~\ref{xs-x3872}. Assuming it comes from the decay of 
a resonance, we use a single Breit-Wigner function to fit the cross section line shape, which yields $M=4200.6^{+7.9}_{-13.3}\pm3.0$~MeV/$c^2$ and $\Gamma=115^{+38}_{-26}\pm12$~MeV. These resonance parameters agree with the $\y$ 
resonance quite well, which strongly hint at the radiative transition process $\y\to\gamma\x$. The connection between two
XYZ particles [$\y$ and $\x$] is established for the first time, which indicate they might share some common nature~\cite{RMP}.

\begin{figure}
\begin{center}
\includegraphics[height=2.0in]{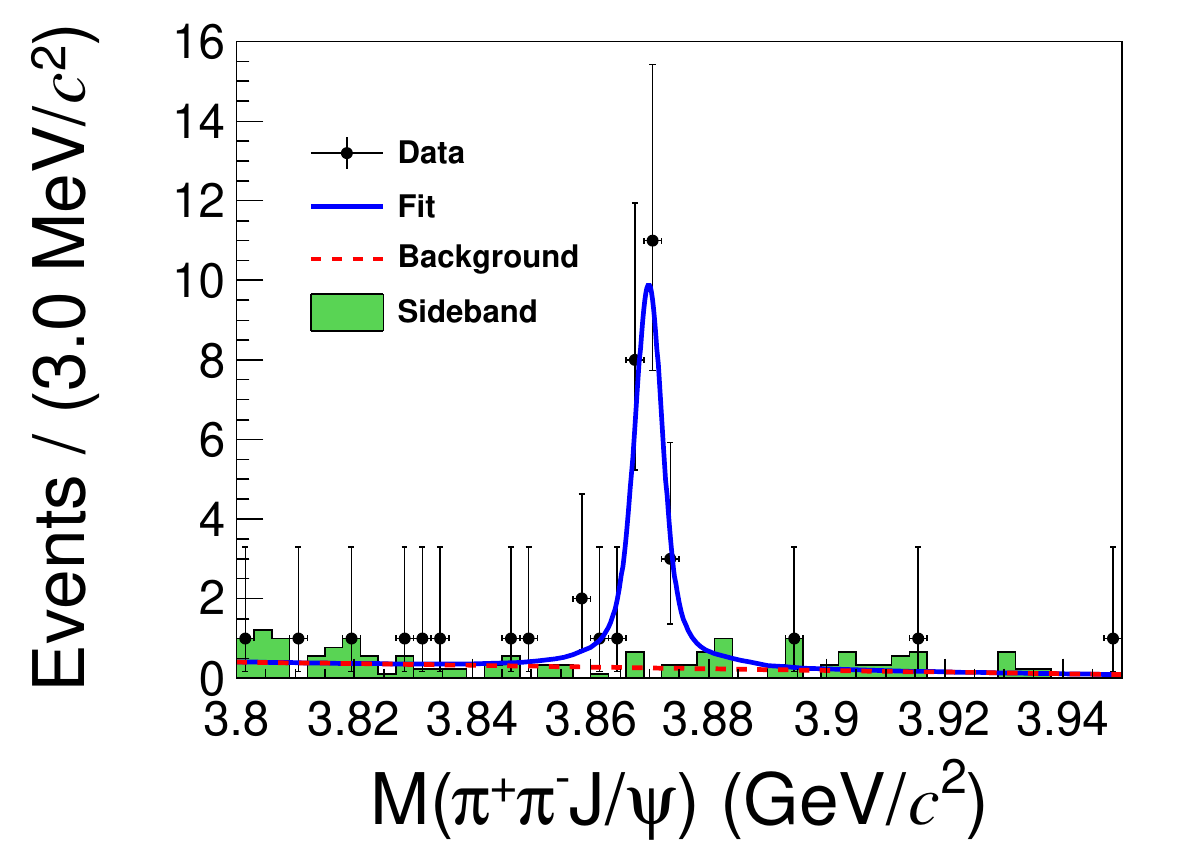}
\includegraphics[height=2.0in]{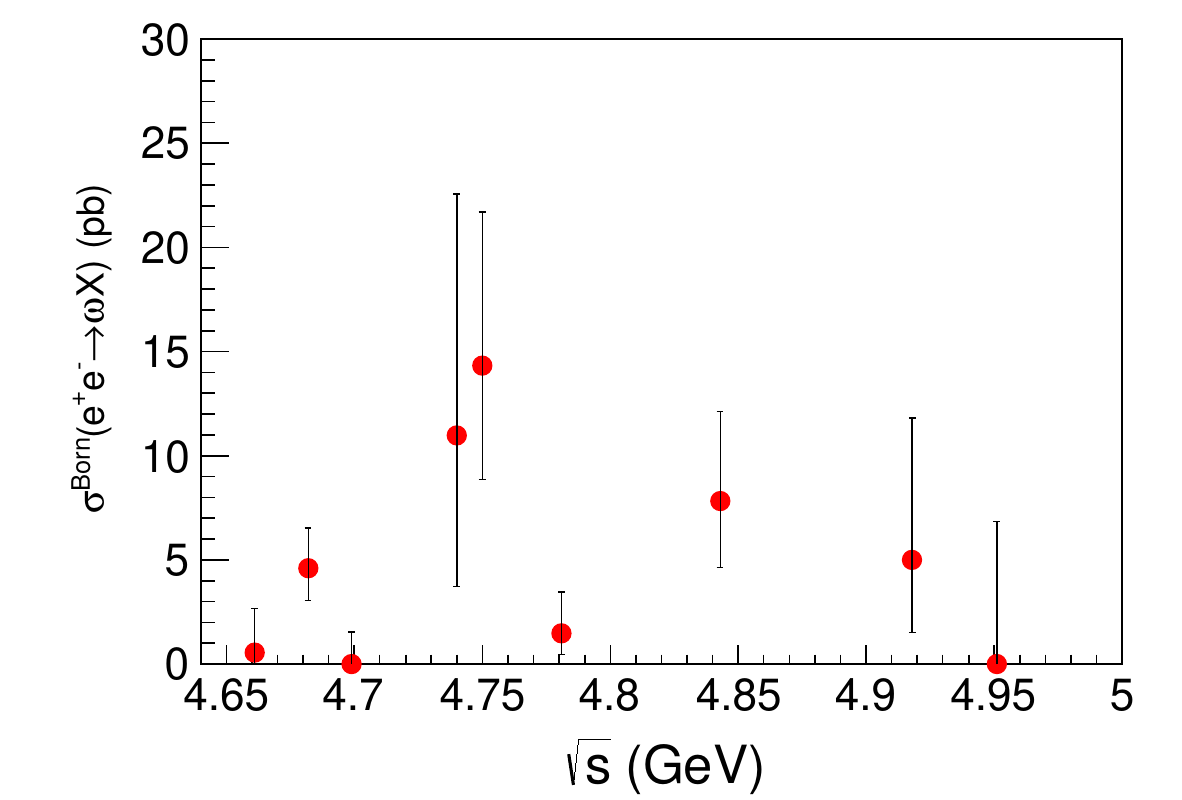}
\caption{The measured $M(\ppjpsi)$ mass distribution (left) and $\sqrt{s}$-dependent production cross section (right) 
in $\EE\to\omega\x$ process at BESIII~\cite{bes3-wx}.}
\label{wx3872}
\end{center}
\end{figure}

In addition to the $\gamma\x$ production, BESIII also investigates other production processes.
According to Vector-Meson-Dominance (VMD), the photon can be replaced with a vector meson, such as the $\EE\to\omega\x$ process. 
With 4.7~fb$^{-1}$ data taken above the $\omega\x$ threshold, BESIII observe
$\EE\to\omega\x$ for the fist time~\cite{bes3-wx}. Figure~\ref{wx3872} (left) shows the 1-dimensional projection plot for $M(\ppjpsi)$, 
where a clear $\x$ signal is observed. A fit to the 1D $M(\ppjpsi)$ distribution 
gives a statistical significance of $>7.8\sigma$. This is a new production mode of $\x$ at BESIII. Although the statistics
is not enough, we try to investigate the $\sqrt{s}$-dependent production cross section of $\EE\to\omega\x$, as shown in 
Fig.~\ref{wx3872} (right). We find the production cross section are significantly higher near $4.75$~GeV, which might be
evidence for a potential resonance in this energy region. With these new $\x$ events, BESIII will further measure the
relative production rate of $\EE\to\gamma\x/\omega\x$, $\omega\x/\omega\chico$, and investigate the production of $\EE\to\phi\x$.

The decays of $\x$ also play an important role in the understanding of its nature. In 2017, BESIII observed the
isospin violation decay $\x\to\piz\chico$, and measured the relative decay rate 
$\frac{\mathcal{B}[\x\to\piz\chico]}{\mathcal{B}[\x\to\ppjpsi]}=0.88^{+0.33}_{-0.27}\pm0.10$~\cite{bes3-pi0chi}. 
The decay of $\x\to\piz\chi_{c0}/\pp\chi_{c0}$ is also possible. Using the radiative production
$\EE\to\gamma\x$ process, the search for $\x\to\piz\chi_{c0}/\pp\chi_{c0}$ is performed in 5 $\chi_{c0}$ decay
channels ($\pp,~\kk,~\pp\pp,~\pp\kk,~\pp\piz\piz$). No significant signal is observed, and an upper limit at 90\%
confidence levels (C.L.) for the relative decay rate is given, which is
$\frac{\mathcal{B}[\x\to\piz\chi_{c0}]}{\mathcal{B}[\x\to\piz\chico]}<4.5$ and
$\frac{\mathcal{B}[\x\to\pp\chi_{c0}]}{\mathcal{B}[\x\to\ppjpsi]}<0.56$~\cite{bes3-pichi}.

\begin{figure}
\begin{center}
\includegraphics[height=1.3in]{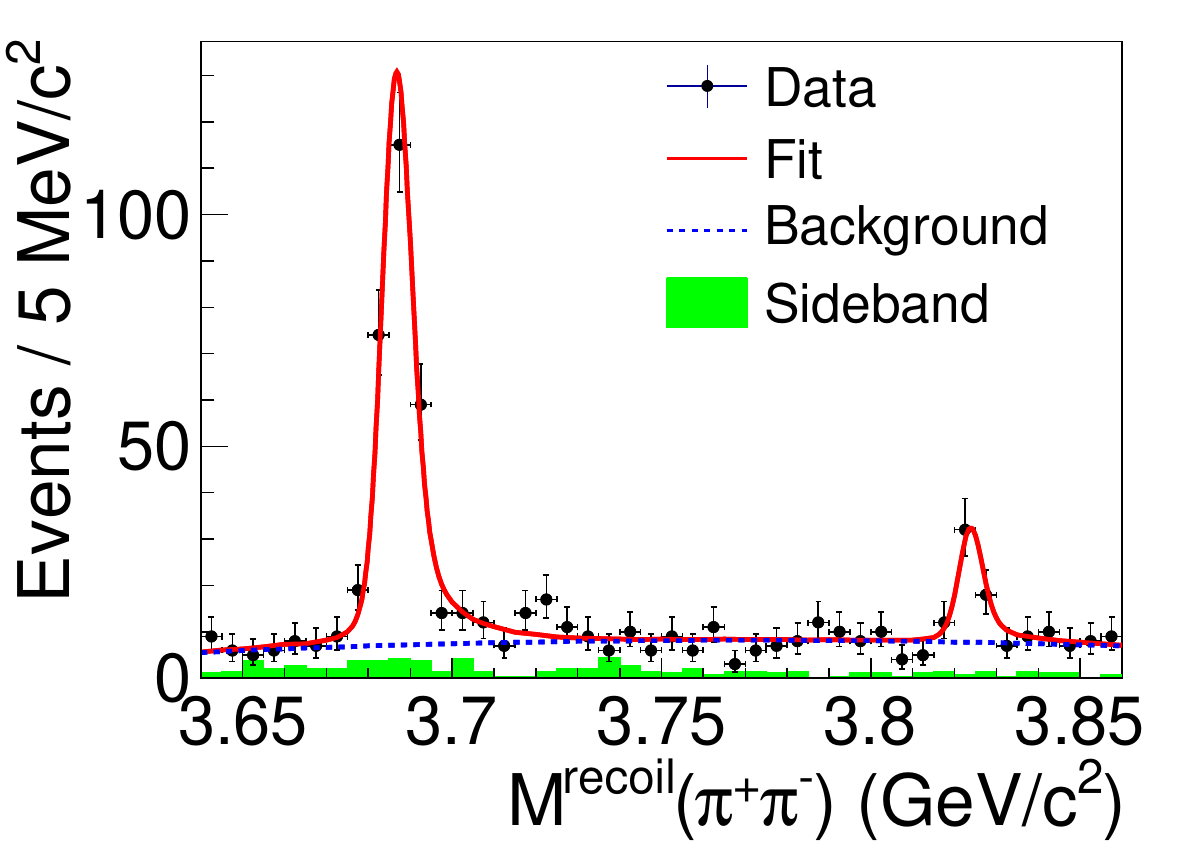}
\includegraphics[height=1.3in]{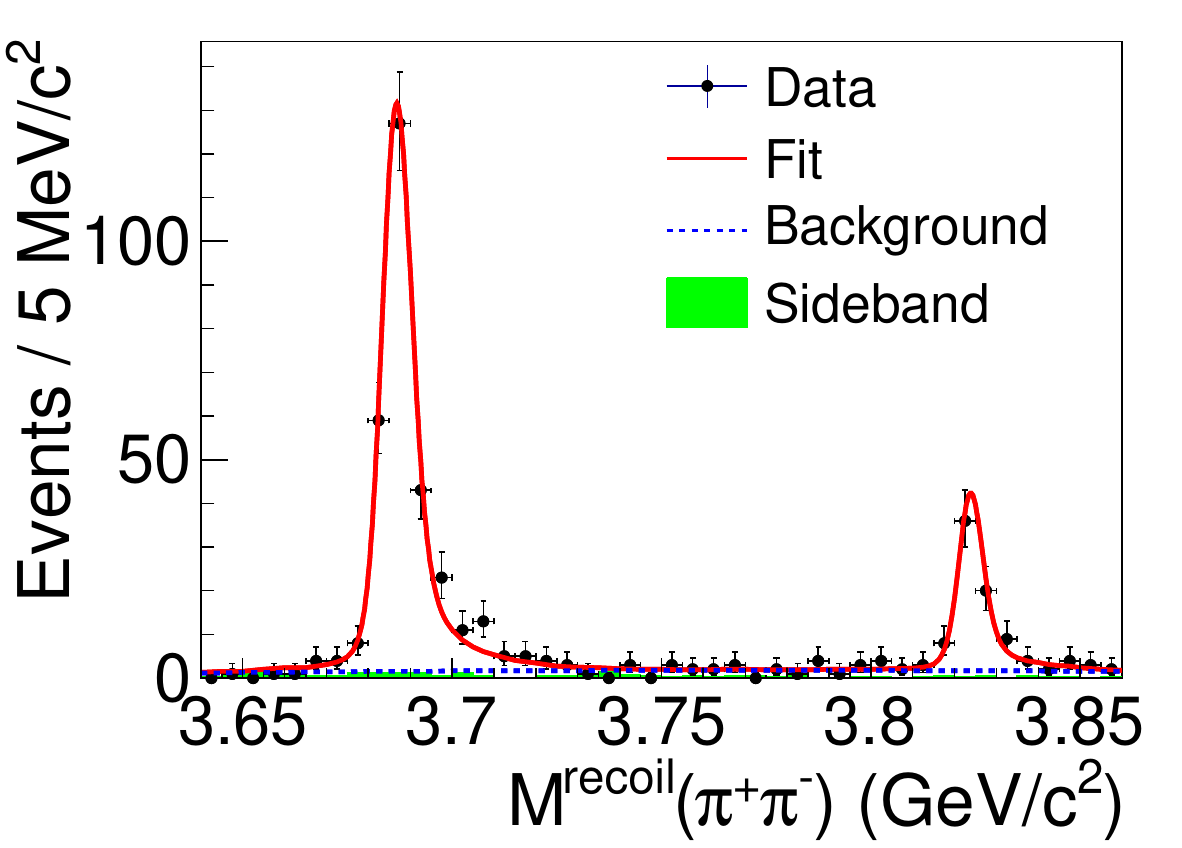}
\includegraphics[height=1.3in]{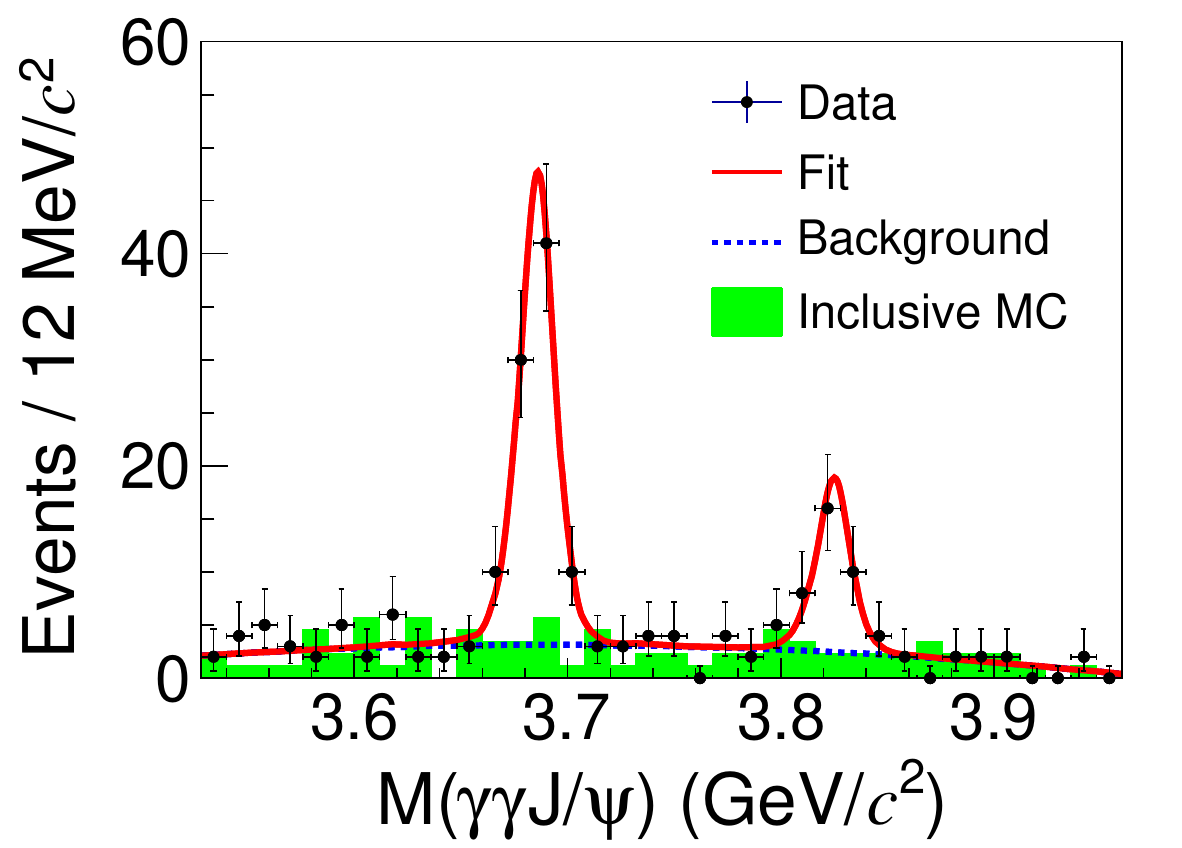}
\caption{The $\psit$ signal observed with partial reconstruction method (left) and full reconstruction method (middle)
in $\EE\to\pp\psit$~\cite{bes3-ppx3823} and in $\EE\to\piz\piz\psit$~\cite{bes3-p0x3823} (right) processes at BESIII.}
\label{psi2}
\end{center}
\end{figure}

The $\psit$ state is well recognized as the $1^3D_2$ charmonium~\cite{pdg}. Using 11.3~fb$^{-1}$ data, the BESIII
experiment precisely study the properties of $\psit$~\cite{bes3-ppx3823}. 
By introducing a partial reconstruction technique, the detection efficiency 
of $\psit$ signal events is increased by a factor of almost two for the $\EE\to\pp\psit$ process. 
The $\psit$ signal is further measured with the
recoil system of low momentum $\pp$ (without kinematic correction), which does not reduce the signal resolution.
Figure~\ref{psi2} shows the $\pp$ recoil mass [$M^{\rm recoil}(\pp)=\sqrt{(P_{\EE}-P_{\pp})^2}$] distribution
both for partially and fully reconstructed events. A 1-dimensional fit to the $M^{\rm recoil}(\pp)$ distribution simultaneously
yields 120 $\psit$ signal events, and give the mass of $\psit$ to be $(3823.12\pm0.43\pm0.13)$~MeV/$c^2$
and width $\Gamma<2.9$~MeV at 90\% C.L. This is the most precise mass measurement and most stringent width constraint
to date. We further measure the isospin neutral process $\EE\to\piz\piz\psit$, and the obtained $M(\GG\jpsi)$ mass
distribution is also shown in Fig.~\ref{psi2} (right). Obvious $\psit$ signal is observed, with a significance of $6\sigma$~\cite{bes3-p0x3823}.
The relative production cross section with respect to the charged mode is measured to be
$\frac{\sigma[\EE\to\piz\piz\psit]}{\sigma[\EE\to\pp\psit]}=0.57\pm0.14$, which agrees well with isospin symmetry (0.5).
The observation of $\EE\to\piz\piz\psit$ process further confirms the $\pp$ system can not originate from $\rho^0$,
which supports the $J^{PC}=2^{--}$ assignment for $\psit$.

% Y
\section{Vector $Y$-states}
\begin{figure}
\begin{center}
\includegraphics[height=2.8in]{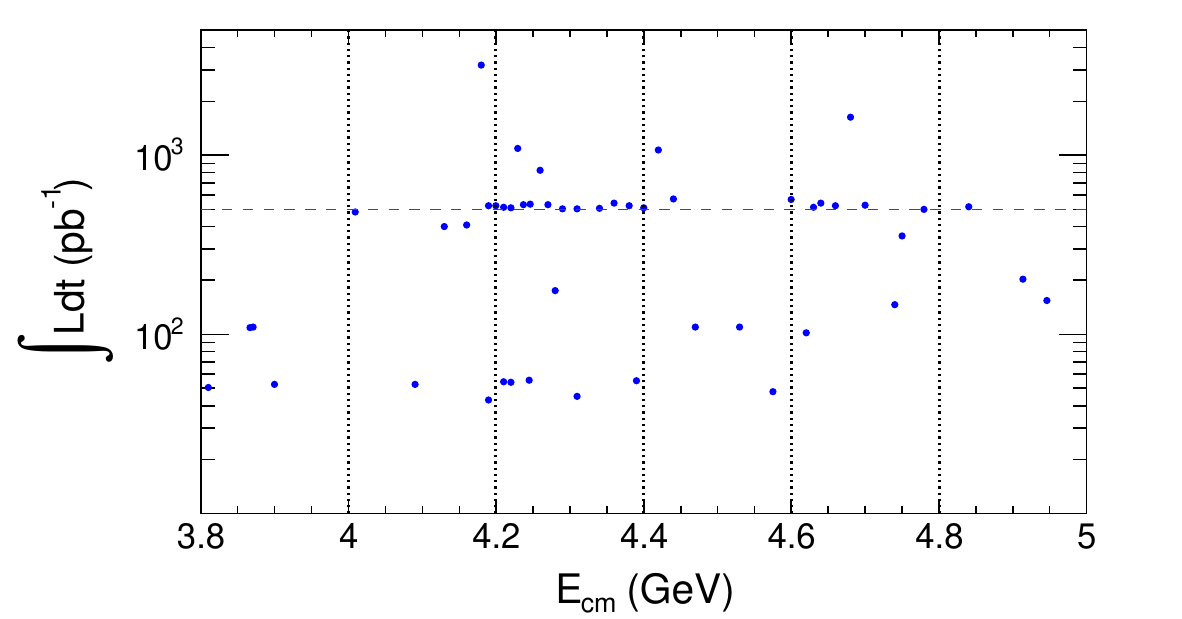}
\caption{The integrated luminosity of the data sets collected by BESIII at various center-of-mass energies. The red dashed horizontal line is equivalent to 500~pb$^{-1}$, which generally represents the luminosity at many center-of-mass energies.}
\label{lum}
\end{center}
\end{figure}

The BESIII experiment runs from $\sqrt{s}=2-5$~GeV. It is a symmetric colliding experiment with luminosity optimized
at $\sqrt{s}=3.773$~GeV. Started from 2013, BESIII did intensive energy scans between 4 and 5~GeV, which allows
precise study of vector $Y$-states in this energy region. Figure~\ref{lum} shows the data set accumulated by BESIII,
with a total integrated luminosity of 21.5~fb$^{-1}$.

\begin{figure}
\begin{center}
\includegraphics[height=3.0in,angle=-90]{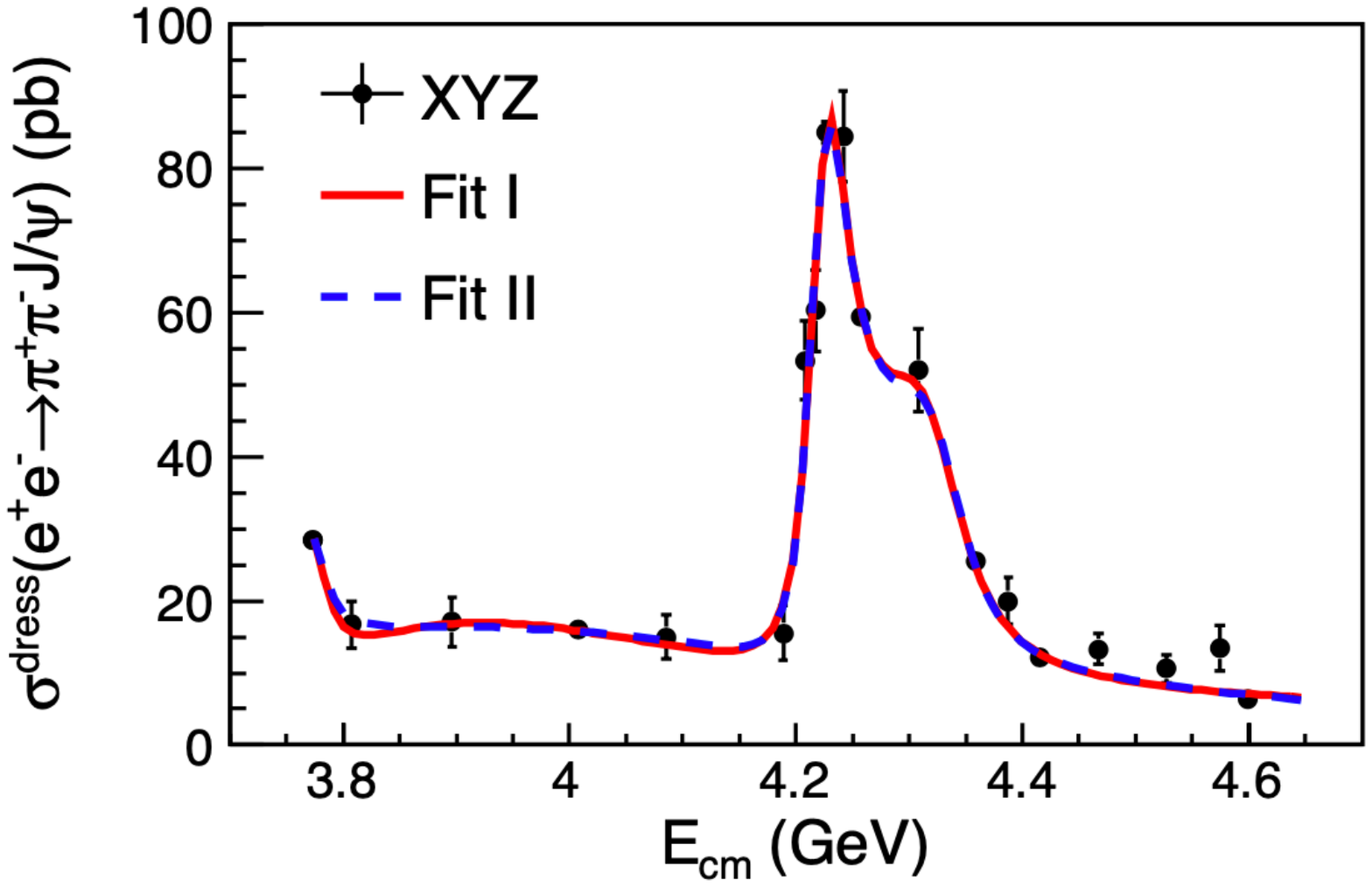}
\caption{The production cross section of $\EE\to\ppjpsi$ measured at BESIII~\cite{bes3-y4260}.}
\label{split}
\end{center}
\end{figure}

With this data set, the $\EE\to\ppjpsi$ cross section was precisely
measured~\cite{bes3-y4260}. We find the $\y$ resonance previously measured by BaBar~\cite{y4260} and Belle~\cite{belle-y4260}
split into two finer substructures, as shown in Fig.~\ref{split}. 
The sharp peak near 4.2~GeV has a mass of $4222.0\pm3.1\pm1.4$~MeV and a width of
$44.1\pm4.3\pm2.0$~MeV. This is the most precise measurement of the $\y$ mass, and it is significantly lower
than previous measurements. A recent update of $\EE\to\ppjpsi$ cross section with more data at BESIII further
confirms the overlap structure of $\y$~\cite{bes3-y4260-2}. Lattice QCD theory calculates the mass of the vector 
hybrid to be 4285~MeV/$c^2$~\cite{hybrid}. The BESIII measurement obviously deviates from this prediction,
and gives new insight about its nature, such as a $DD_1$ molecule.

\begin{figure}
\begin{center}
\includegraphics[height=2.9in,angle=-90]{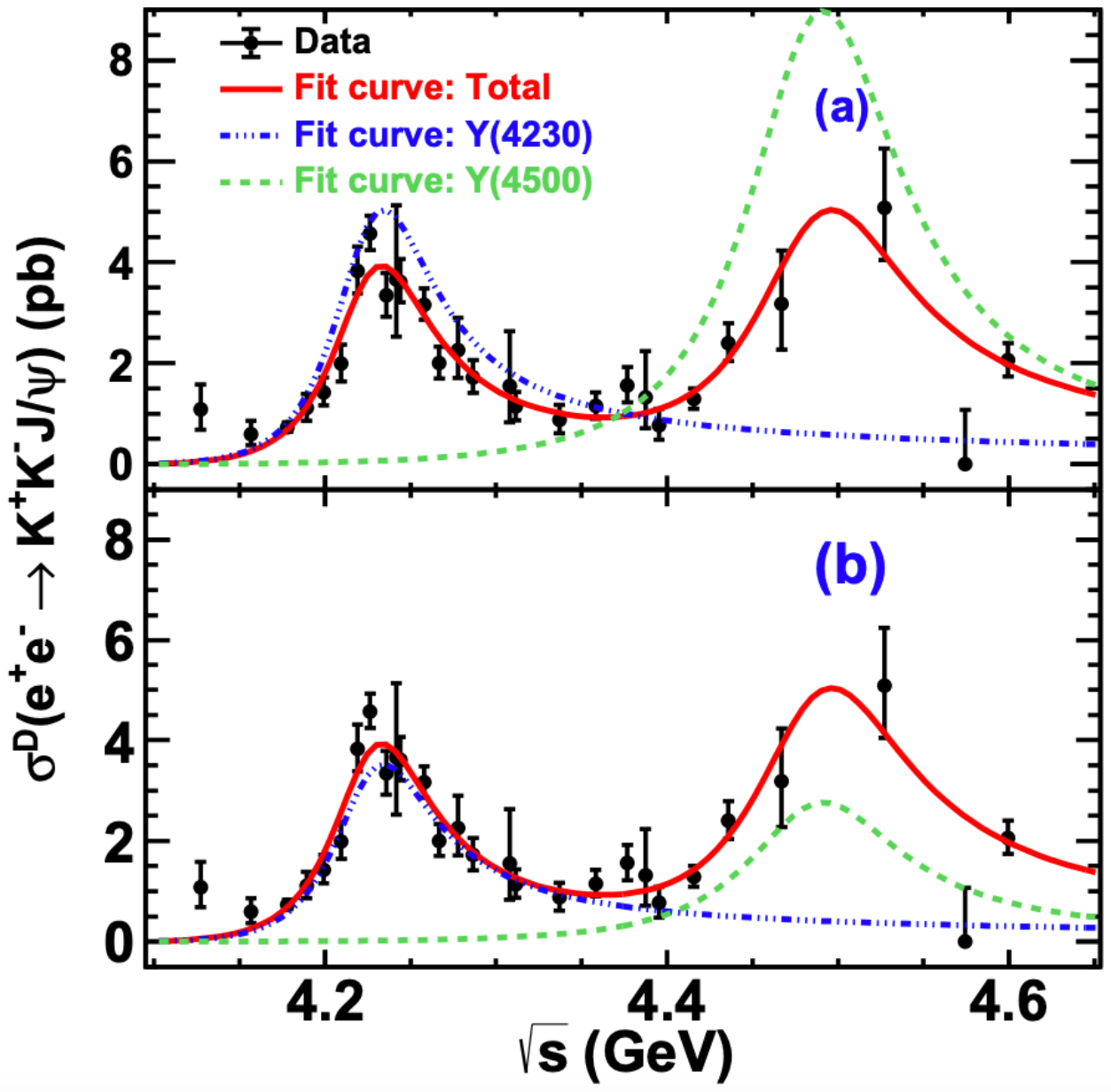}
\includegraphics[height=2.9in,angle=-90]{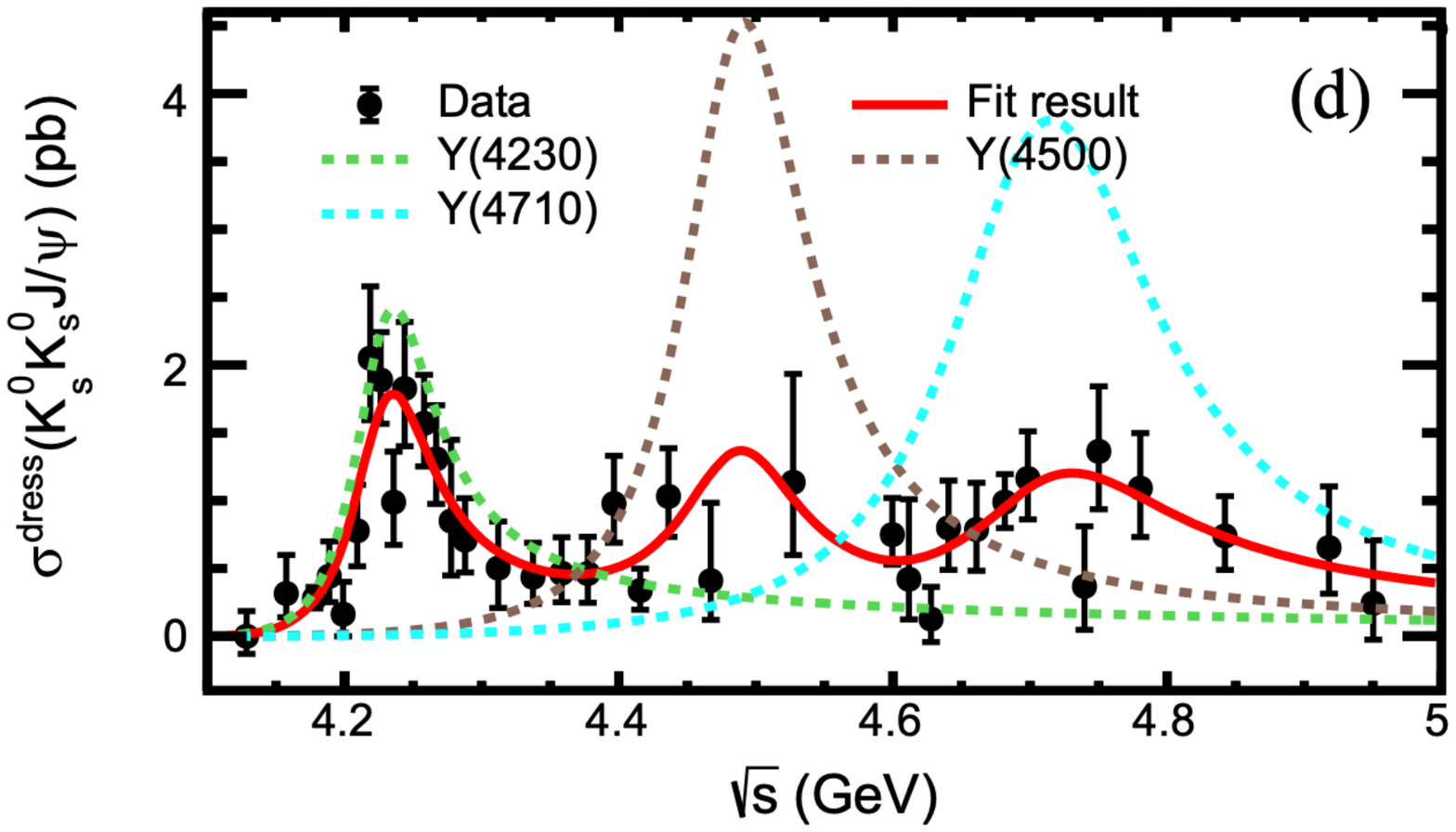}
\caption{The production cross section of $\EE\to\kk\jpsi$~\cite{bes3-y4500} (left) and $K_sK_s\jpsi$~\cite{bes3-y4710} (right) at BESIII.}
\label{kkjpsi}
\end{center}
\end{figure}

The $\EE\to\kk\jpsi$ and $K_sK_s\jpsi$ cross section were measured by BESIII from threshold up to 4.95~GeV.
In the $\EE\to\kk\jpsi$ study, we firstly measure the production cross section between 4.13 and 4.60~GeV~\cite{bes3-y4500}
and an update study with extended center-of-mass energy was already ongoing.
Figure~\ref{kkjpsi} (left) shows the measured $\kk\jpsi$ cross section.
Two resonance structures are observed. The lower one is consistent with the $\y$ measured by BESIII~\cite{bes3-y4260,bes3-y4260-2},
which is referred as $Y(4230)$ now by PDG~\cite{pdg}. A new structure, denoted as $Y(4500)$ 
with mass $4484.7\pm13.3\pm24.1$~MeV/$c^2$,
and width $111.1\pm30.1\pm15.2$~MeV was observed for the first time, with a statistical significance $>8\sigma$.
In the $K_sK_s\jpsi$ process, cross section measurement was even extended to 4.95~GeV, as shown in Fig.~\ref{kkjpsi} (right). In addition
to the $Y(4230)$ and $Y(4500)$ observed in the charged mode, a higher mass structure, 
$Y(4710)$ was evident~\cite{bes3-y4710}. Further confirmation of this resonance in the charged mode is important
and necessary.

\begin{figure}
\begin{center}
\includegraphics[height=1.38in]{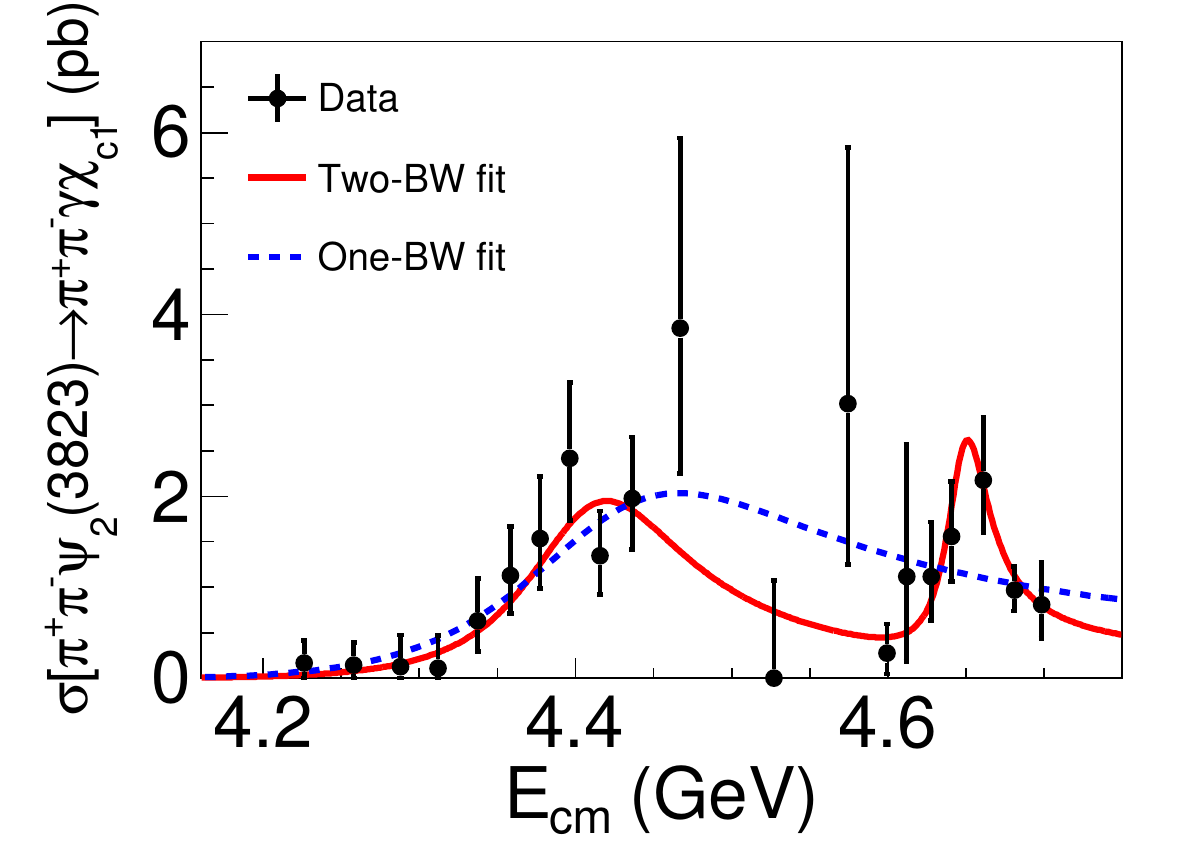}
\includegraphics[height=1.4in]{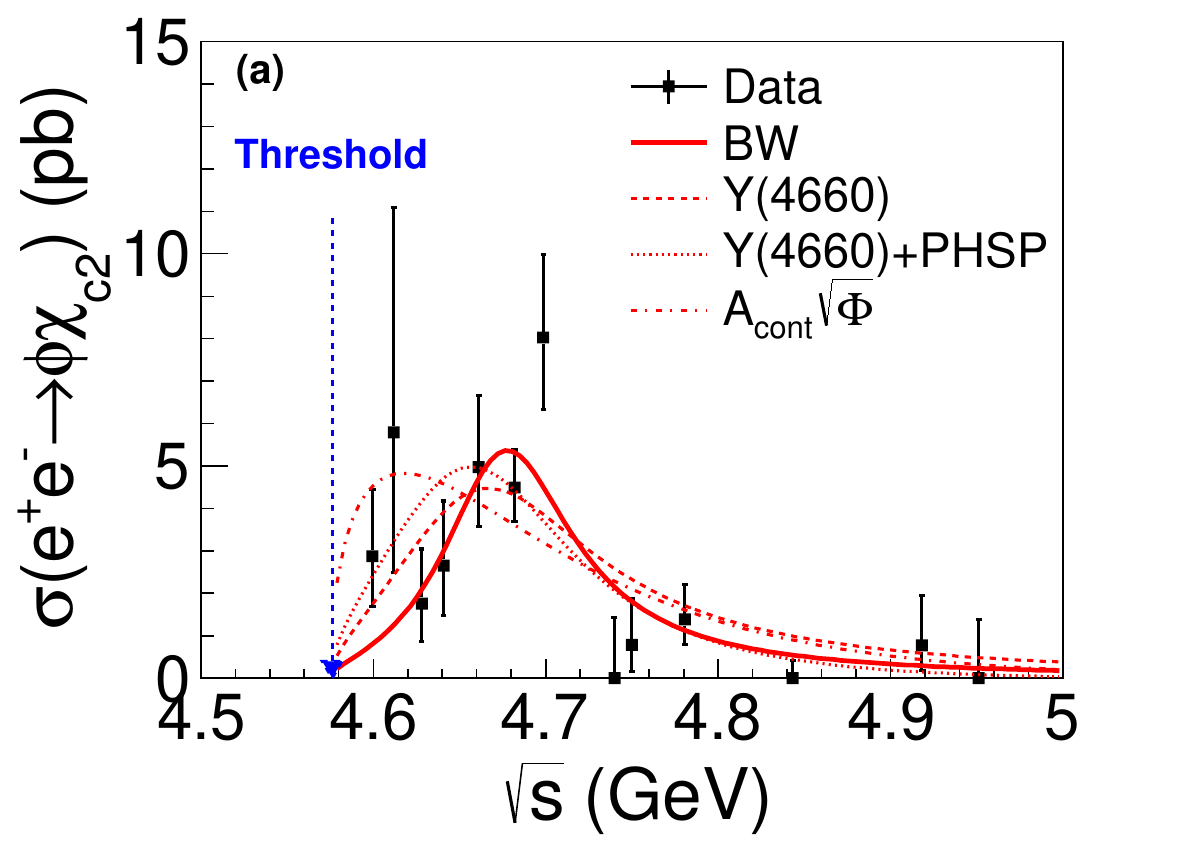}
\includegraphics[height=1.4in]{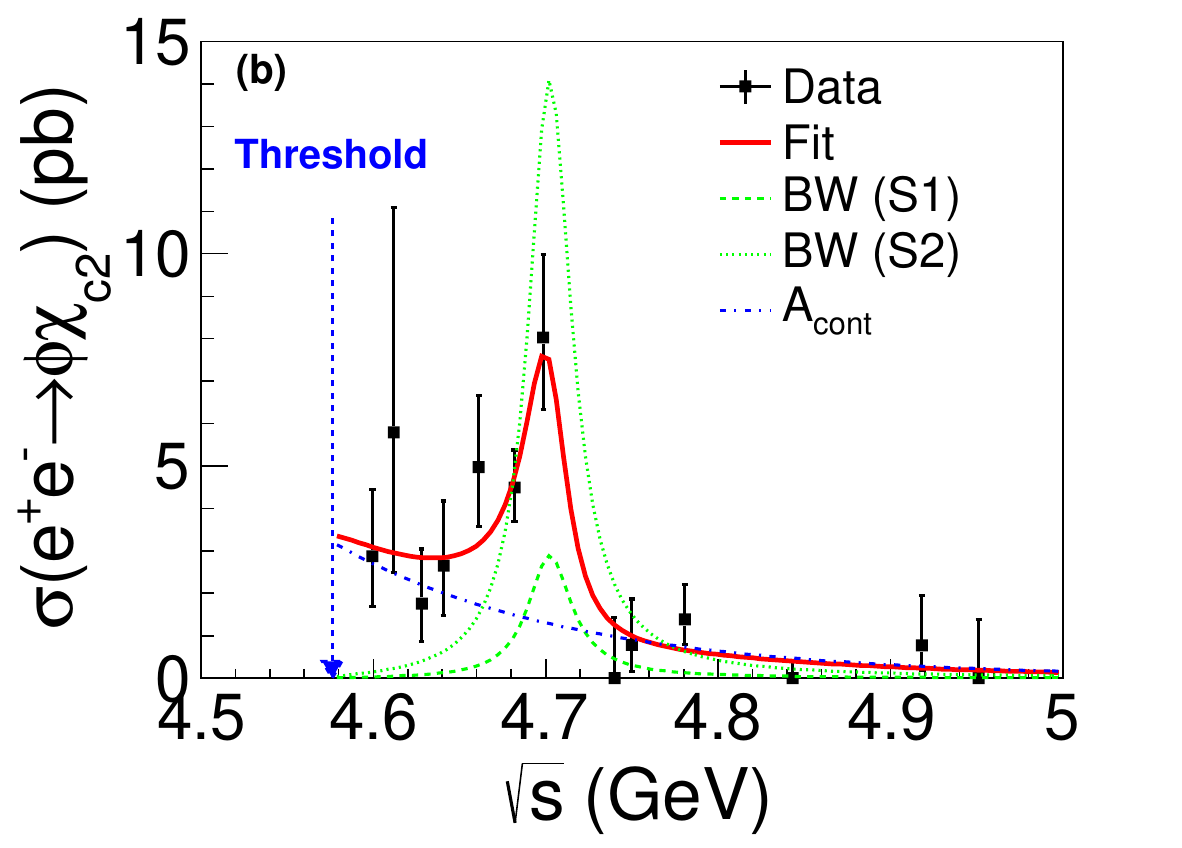}
\caption{The production cross section of $\EE\to\pp\psit$~\cite{bes3-ppx3823} (left) 
and $\phi\chict$~\cite{bes3-fchi2} (middle and right) at BESIII.}
\label{ppx3823}
\end{center}
\end{figure}

The $\EE\to\pp\psit$ cross section was investigated from 4.20 to 4.70~GeV by BESIII~\cite{bes3-ppx3823}, and we
find that a resonance structure is needed to describe the cross section line shape. 
Figure~\ref{ppx3823} (left) shows the measured $\EE\to\pp\psit$ cross section.
Two fit models are 
used to extract the resonance parameters. One with two coherent Breit-Wigner resonances is favored, 
and it yields two resonance structures which agree with the $Y(4360)$ and $Y(4660)$ resonances
observed in $\EE\to\pp\psip$ process by B-factories before~\cite{pdg}. If this is the case, $\pp\psit$ is the
second decay channel of the $Y(4660)$, and the relative decay width 
$\frac{\mathcal{B}[Y(4660)\to\pp\psit]}{\mathcal{B}[Y(4660)\to\pp\psip]}\sim20\%$ is sizable.
This large decay rate give important information regarding its nature.
%differs from theoretical interpretations of the $Y(4660)$, such as a $f_0(980)\psip$ molecule~\cite{} etc.
A single resonance model is also used to fit the cross section, although we are not able to rule out
it due to limited statistics.
%with mass about 4417~MeV and width 250~MeV is also possible to fit the cross section,

The $\EE\to\phi\chi_{c1,c2}$ processes were firstly observed by BESIII at $\sqrt{s}=4.60$~GeV~\cite{bes3-fchi}. With more data
(6.4~fb$^{-1}$) from 4.60 to 4.95~GeV, obvious $\phi\chi_{c1,c2}$ production events (significance $>10\sigma$) were observed,
and the $\sqrt{s}$-dependent production cross section were also measured~\cite{bes3-fchi2}. In the $\phi\chico$ process,
a continuum amplitude is able to describe the cross section line shape. While for the $\phi\chict$ process,
resonance structures are evident in the cross section line shape. Two fit models are used to extract
the resonance parameters. One with a single free Breit-Wigner function yields
a resonance with parameters consistent with the $Y(4660)$, as shown in Fig.~\ref{ppx3823} (middle). 
The other with a coherent sum of 
a Breit-Wigner resonance and a continuum amplitude yields a new structure with mass around 4.70~GeV,
as shown in Fig.~\ref{ppx3823} (right).
The statistical significance of these structures are at $3\sigma$ level due to limited statistics,
and we are not able to distinguish these two fit models at present. In addition, possible $\phi\jpsi$
resonances, the $X(4274)$, $X(4500)$, and $X(4700)$~\cite{lhcb-X} were also searched in $\EE\to\gamma X$.
No obvious signal is observed, and upper limit at 90\% C.L. are given for the production cross section.

\begin{figure}
\begin{center}
\includegraphics[height=2.76in,angle=-90]{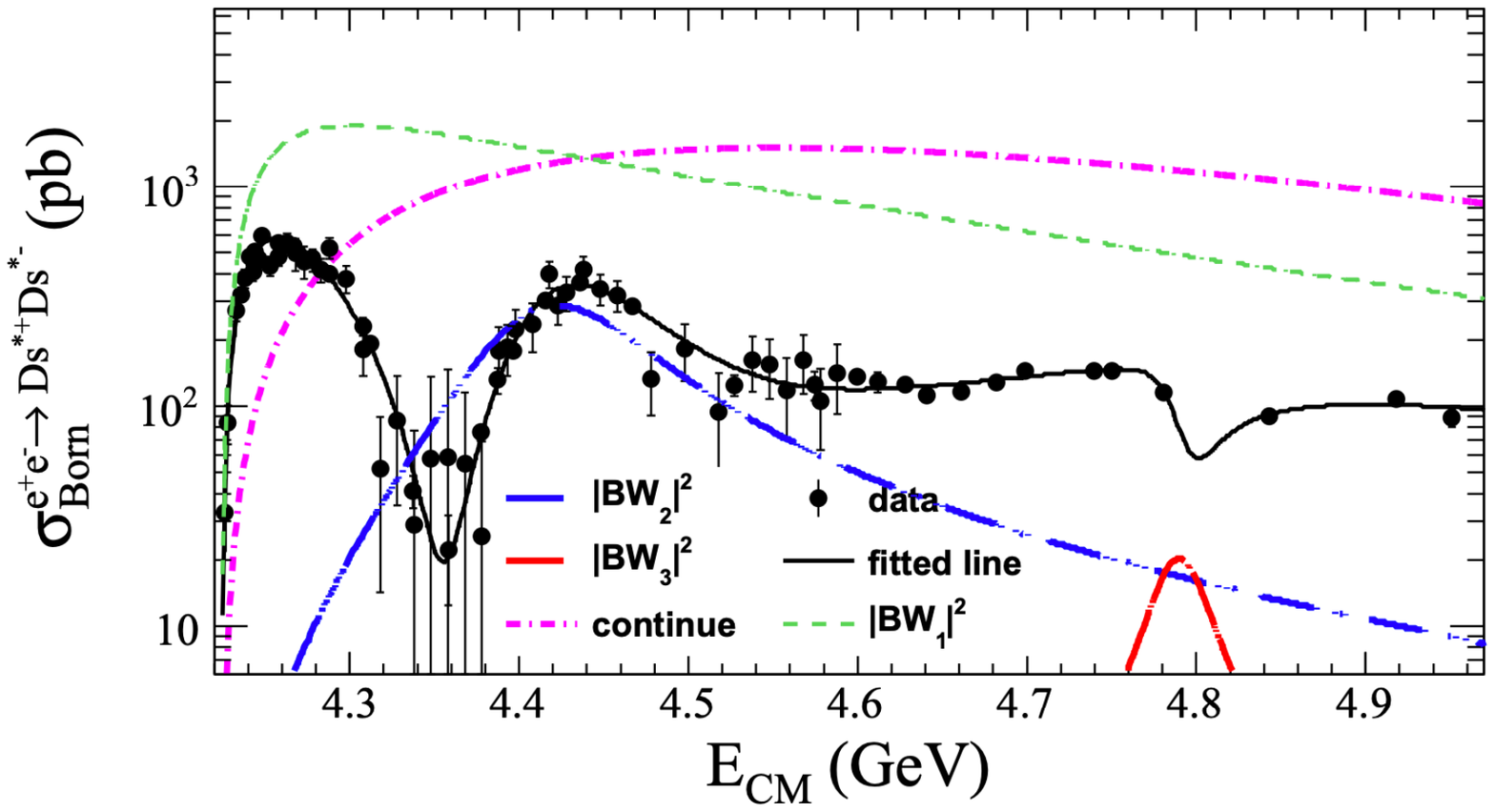}
\includegraphics[height=2.8in,angle=-90]{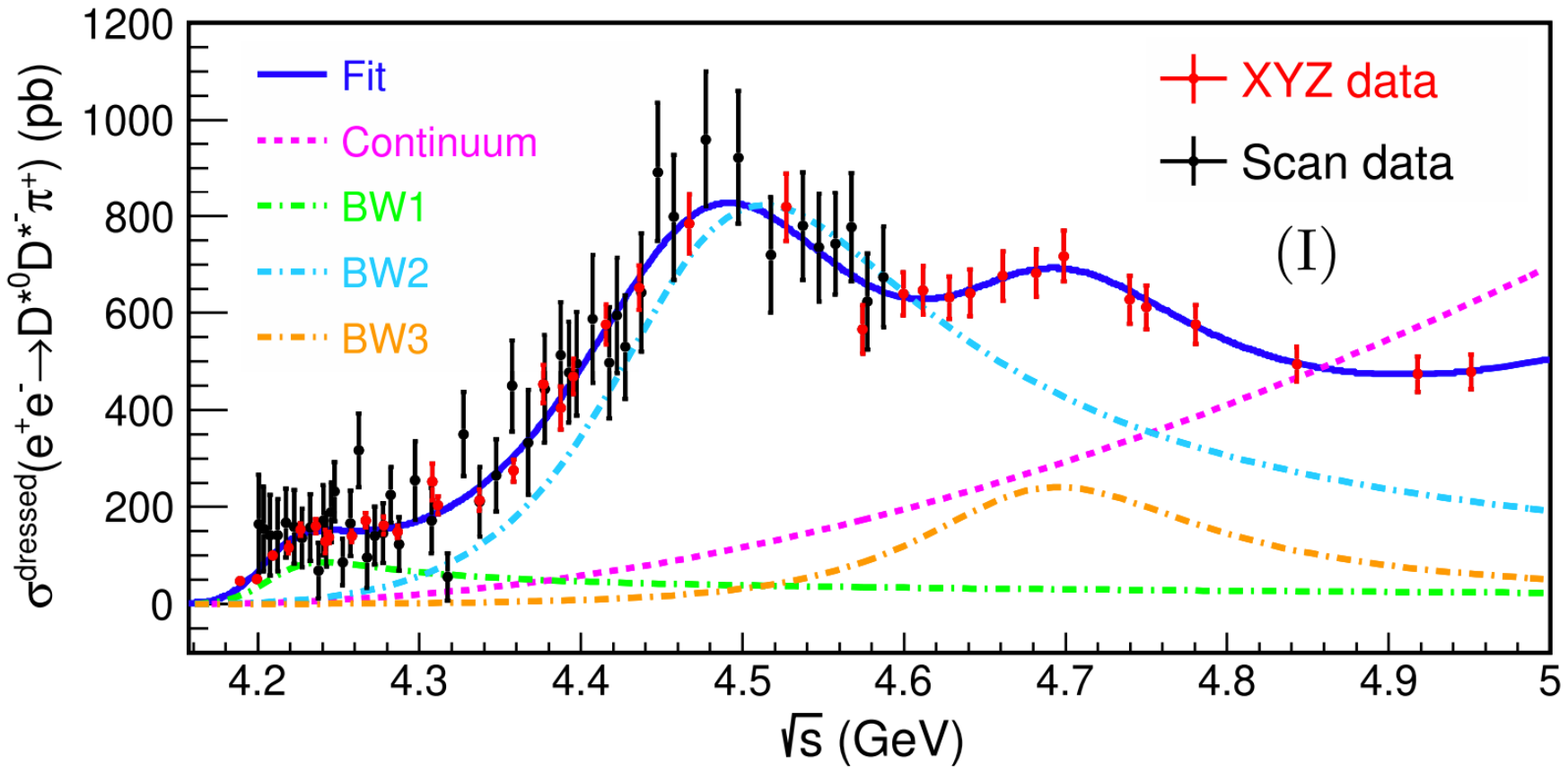}
\caption{The production cross section of $\EE\to D_s^{*+}D_s^{*-}$~\cite{dsds} (left) and $D^{*0}D^{*-}\pip$~\cite{ddpi} (right) at BESIII.}
\label{opencharm}
\end{center}
\end{figure}

The open-charm production process $\EE\to D_s^{*+}D_s^{*-}$
was precisely measured recently~\cite{dsds}, and is shown in Fig.~\ref{opencharm} (left). 
In the $D_s^{*+}D_s^{*-}$ process, the cross section line shape is non-trivial.
The coherent sum of three Breit-Wigner resonances and a continuum amplitude is used to fit the
cross section line shape. The parameter of the first resonance is consistent with the $\psi(4160)$ or the $Y(4230)$~\cite{pdg}.
For the $Y(4230)$ case, it means the open-charm coupling of $Y(4230)$ is much larger than $\ppjpsi$, and thus
alter the interpretation of this resonance. The second resonance with a mass consistent with the $\psi(4415)$ 
is observed in $D_s^{*+}D_s^{*-}$ for the fist time. It also shows a dip around 4.8~GeV in the cross section
line shape. To account for this, a third structure with mass around $4790$~MeV is introduced and necessary to 
improve the fit quality. Its statistical significance is estimated to be $>6\sigma$.

Another open-charm production process $\EE\to D^{*0}D^{*-}\pip$ was also measured~\cite{ddpi} and is shown
in Fig.~\ref{opencharm} (right). In order to
model the non-trivial cross section line shape, again the coherent sum of three Breit-Wigner resonances and a continuum
amplitude is used. The fit yield three resonance structures. The first one is consistent with the $Y(4230)$ and shows
a large coupling to open-charm final state. The second one is consistent with the $Y(4500)$~\cite{bes3-y4500} 
observed in $\kk\jpsi$ process
before, but shows a much larger coupling than $\kk\jpsi$. The third resonance is consistent with the $Y(4660)$ resonance, 
and it is the first time to observe open-charm decay of this resonance.

It should be noted that in order to describe these cross sections, simple models with the coherent sum
of Breit-Wigner resonances are used at the moment. It is not perfect, and more advanced methods,
such as $K$-matrix can be used in future to fully understand these vector resonances.

\begin{figure}
\begin{center}
\includegraphics[height=4.6in,angle=-90]{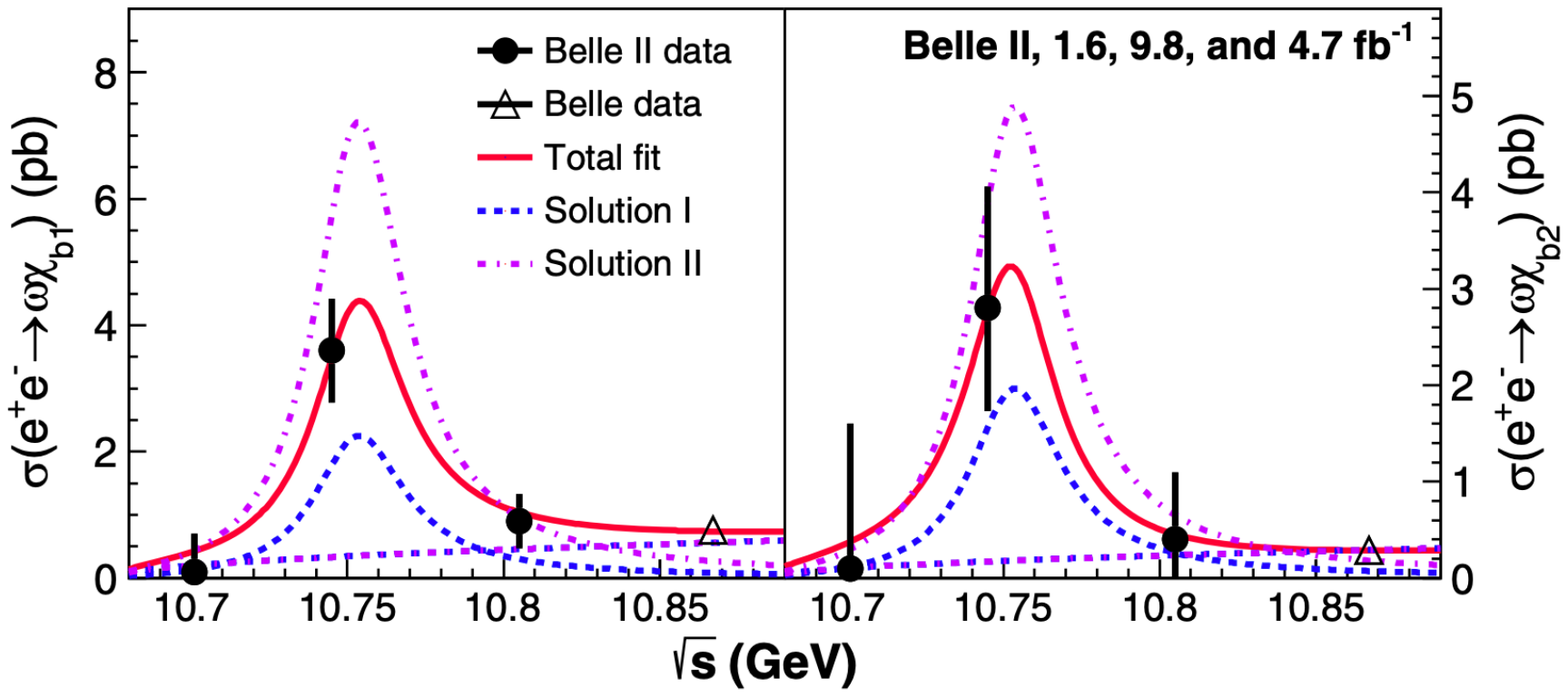}
\caption{The production cross section of $\EE\to\omega\chi_{b1}$ (left) and $\omega\chi_{b2}$ (right) at Belle II~\cite{chib}.}
\label{bottom}
\end{center}
\end{figure}

In the bottom sector, the Belle II experiment recently did an energy scan around 10.75~GeV. With this data set,
the $\EE\to\omega\chi_{b1,b2}$ process was significantly observed~\cite{chib}. 
Combined with a previous data set collected by Belle at $\sqrt{s}=10.86$~GeV, Belle II measured the energy
dependent cross section of $\EE\to\omega\chi_{b1,b2}$ as shown in Fig.~\ref{bottom}. 
The production cross section is significantly larger at
$\sqrt{s}=10.75$~GeV. Assuming it comes from a single resonance and by fixing its parameters to the $Y(10750)$
observed before, Belle II measured the relative rate 
$\frac{\mathcal{B}[Y(10750)\to\omega\chi_{b1}]}{\mathcal{B}[Y(10750)\to\omega\chi_{b2}]}=1.3\pm0.6$.
This value differs from a pure D-wave prediction for the $Y(10750)$~\cite{d-y10750}.

\section{Charged $Z_{cs}$ states}
\begin{figure}
\begin{center}
\includegraphics[height=2.8in,angle=-90]{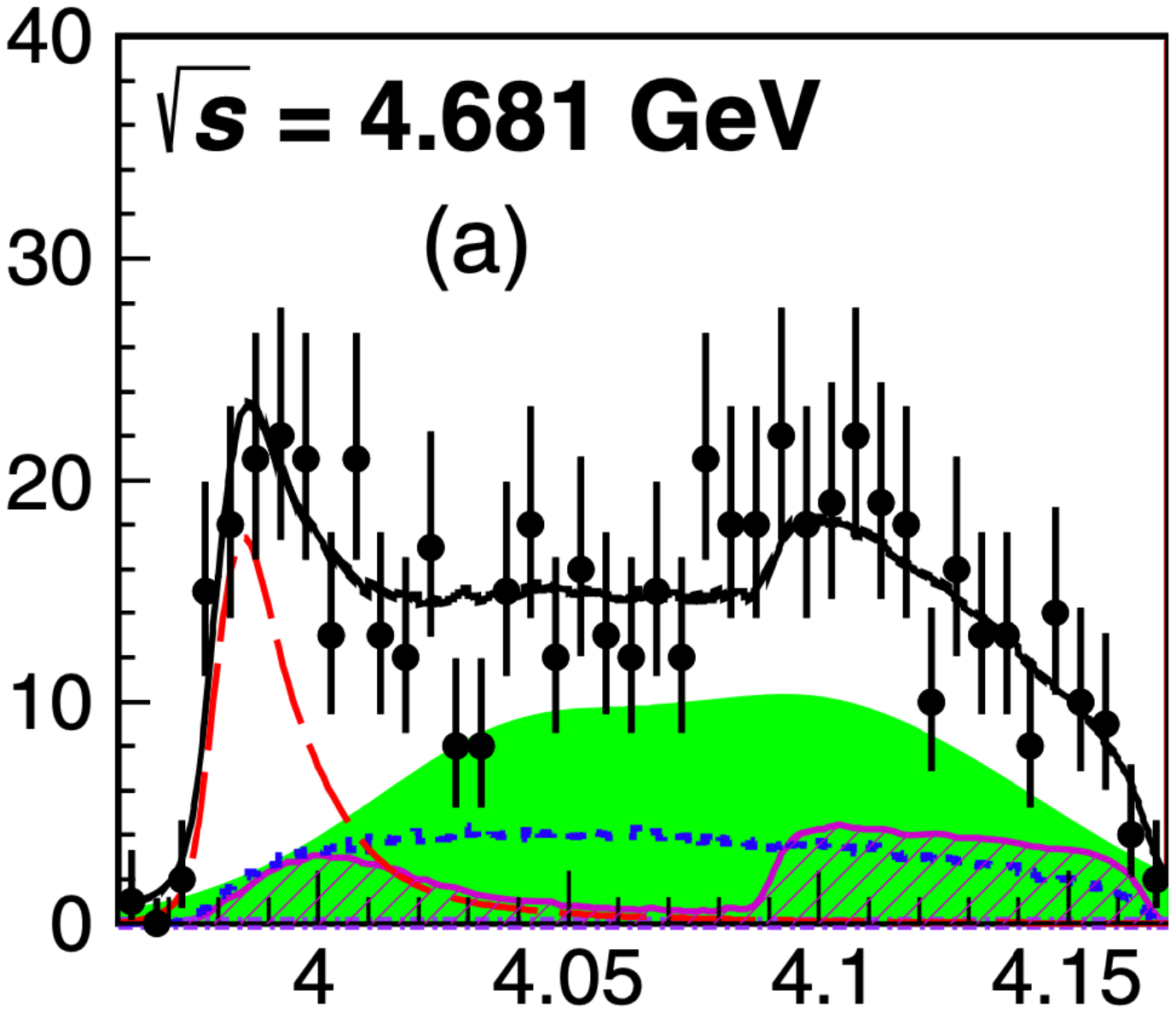}
\includegraphics[height=2.8in,angle=-90]{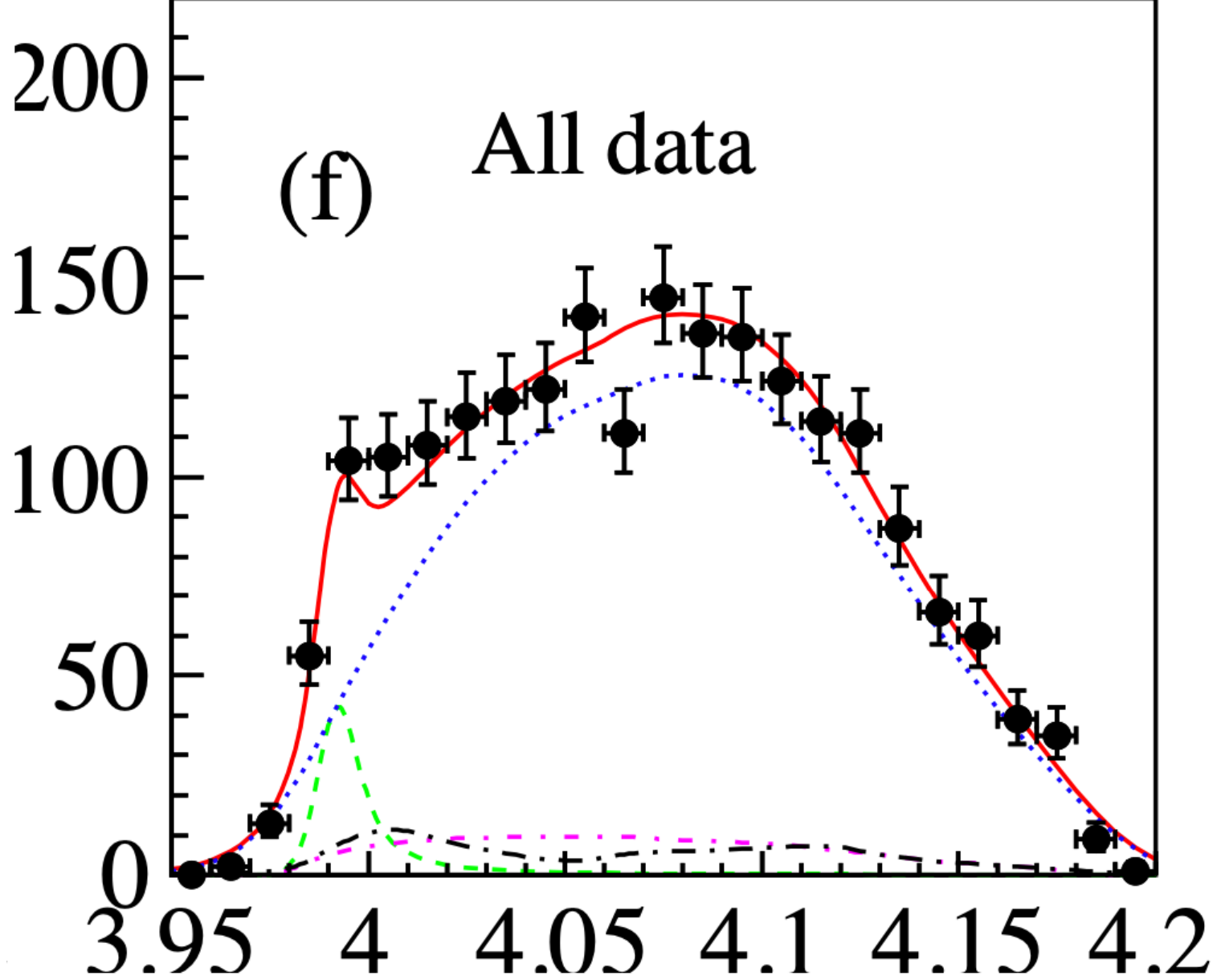}
\caption{The measured $M(D_sD^*/D_s^*D)^\pm$ events distribution per 5~MeV/$c^2$~\cite{bes3-zcs} (left) 
and $M(D_sD^*/D_s^*D)^0$ events distribution per 10~MeV/$c^2$~\cite{bes3-zcs-neu} (right) at BESIII.}
\label{zcs}
\end{center}
\end{figure}

The charged $Z_c$-state is a smoking gun for the existence of exotic hadrons as it carries electric charge. In 2013, the BESIII experiment discovered 
the charged $\z$~\cite{zc3900}, which is widely considered the first confirmed four-quark particle~\cite{fourquark}.
Since the mass of $\z$ is near the $DD^*$ threshold, $\z$ could be a hadronic molecule. In the following
years, BESIII further study other charm meson pair system. With the 3.7~fb$^{-1}$ data taken
between 4.60 and 4.70~GeV, BESIII studied the $\EE\to K^+(D_sD^*/D_s^*D)^-$ process~\cite{bes3-zcs}.
A near threshold enhancement was observed in the $(D_sD^*/D_s^*D)^\pm$ mass distribution,
as shown in Fig.~\ref{zcs} (left). Using a Breit-Wigner
function to fit the mass distribution yields a resonance with a mass $3982.5^{+1.8}_{-2.6}\pm2.1$~MeV/$c^2$
and a width $12.8^{+5.3}_{-4.4}\pm3.0$~MeV. This structure is referred as $Z_{cs}(3985)$ as it contains a
strange quark. If this is the case, the $Z_{cs}(3985)$ could be a $SU(3)$ partner of the $\z$.
Using the same data set, the $\EE\to K_s(D_sD^*/D_s^*D)^0$ was also analyzed~\cite{bes3-zcs-neu}. A neutral structure
was evident in the $(D_sD^*/D_s^*D)^0$ system with a statistical significance of $4.6\sigma$,
as shown in Fig.~\ref{zcs} (right).
This neutral structure could be an isospin partner of the charged $Z_{cs}(3985)$.

\section{Summary}
With the running of the BESIII experiment in China and the Belle II experiment in Japan,
big progresses have been made in the heavy exotic hadron spectroscopy at $\EE$ colliders in recent years.
The $\x$ studies have entered into a precision stage, with precise measurements of its properties as well as 
precise line shape measurement. There are many new vector $Y$ structures observed in the charmonium
and bottomonium energy region. A multi-channel analysis will help us understand their nature in a better way.
BESIII also observed the new charged $Z_{cs}$ state, which could be the $SU(3)$ partner of $\z$.
The XYZ particles zoo becomes much bigger now.

For the near future, two projects will be launched and be the major players in this field. 
The BESIII experiment is going to be upgraded, with the luminosity optimized between 3.77 and 4.7~GeV 
and $\EE$ center-of-mass energy extended to 5.6~GeV. The Belle II experiment will 
collect 50~ab$^{-1}$ data finally, which also allows a rich physics program for hadron spectroscopy.

\end{document}